# Modern success in channeling study and applications at the U-70 accelerator of IHEP

Yu.A. Chesnokov, A.G. Afonin, V.T. Baranov, G.I. Britvich,
P.N. Chirkov, V.A. Maisheev, V.I. Terekhov, I.A. Yazynin

*Institute for High Energy Physics, Protvino, 142281, Russia*

**Summary**. – The report presents an overview of the results obtained at U-70 accelerator of IHEP to use bent crystals for beam control of high energy protons. Considerable attention is paid to practical application of crystals to create new modes of beam extraction from the accelerator to ensure experiments on high energy physics. It was shown that with the crystal deflectors the efficiency reached ~ 90% with intensity up to $10^{12}$ protons per cycle of U-70. The results of experiments on the use of crystals to enhance the effectiveness of the absorption of the unused beam, as well as the use of crystals for collimation of beam halo are presented. Perspectives to use of bent crystals to extract low energy light ions from U-70 are also discussed.

## 1. – Introduction

Ideas of use the particle channeling in bent crystals for steer the beams have been checked up and advanced in many experiments (see [1, 2, 3] and references herein). This method has found the widest practical application in U-70 accelerator of IHEP, where crystals are used in regular runs for beam extraction and forming [4,5].

## 2. – Beam extraction from U-70 ring by means of bent crystals

Different types of extraction schemes were realized by bent crystal. In first case high efficiency of extraction up to 85% is reached applying short silicon crystals Si 19, 22, 106 (Fig. 1). Short crystals with 2 mm in length and about 1 mrad bend take a role of first septum, than additional magnets provided deflection of circulated beam out of the ring. Such high efficiency is based on multi-turn process when particles which are not captured into channeling mode on the first passage of crystal can be efficiently captured on next encounters with crystal. Extraction efficiency of 70 GeV protons versus crystal length is presented in Fig. 2 in comparison with simulation.

On Fig 3 the crystal extraction efficiency was measured as a function of proton energy. There is good agreement between the measured and calculated efficiency of bent crystal. The reduction of efficiency with decreasing of energy is explained by the increase of multiple scattering angle and the decrease in the dechanneling length. The obtained dependence shows also that by using the same crystal one can extract beams in a broad energy range (40÷70) GeV with efficiency above 60%.

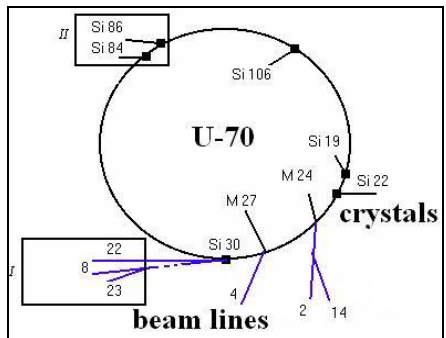

Fig. 1: Crystal location at U-70 ring.

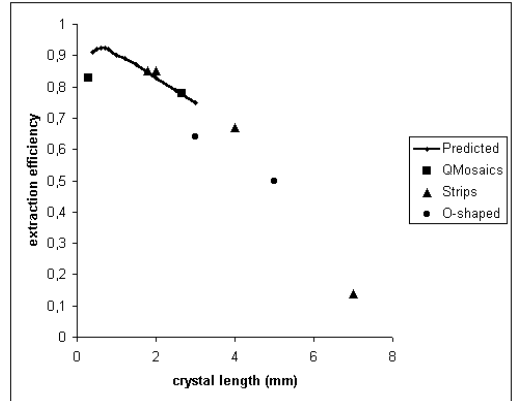

Fig. 2: Extraction efficiency dependence versus crystal length.

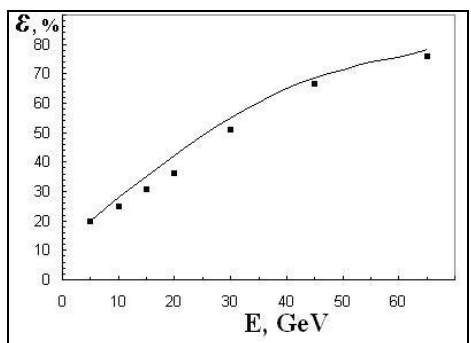

Fig. 3: Crystal extraction efficiency versus proton energy.

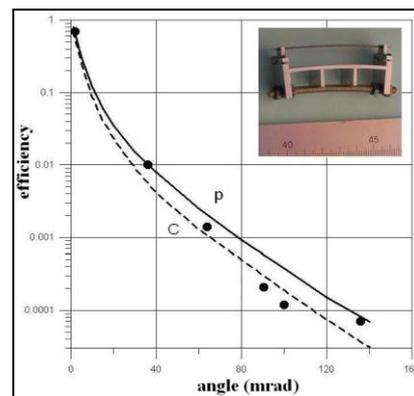

Fig. 4: Extraction efficiency dependence versus crystal bend angle.

Important point is the crystal radiation hardness at accelerators. The limit of irradiation that channeling crystal survives was obtained in CERN and BNL experiments: ~$2\times10^{20}$ proton per $cm^2$. Our experiments confirm these results. Crystals don't loose channeling properties during two runs 1400 hours each. As far as heat loads are concerned, our experience shows that crystal with efficiency (80÷85)% ensures beam extraction at intensity up to $10^{12}$ particle/cycle and duration (1÷2) s per cycle, suiting the requirements of the most experiments at IHEP accelerator. The created extraction method works at IHEP since the end of 1999 in every run of accelerator. Detailed consideration of all the systems of this extraction is given in [5].

Other option of extraction using long crystals (few cm in length, few tens mrad bend) was investigated. Efficiency of extraction dropped with increasing of deflection angle (Fig.4), but these moderate intensity beams are also promising for providing of physical program with protons and ions in few IHEP beamlines [6].

## 2. – Use of crystal channeling to improve beam collimation in U-70.

The classic two-stage collimation system for loss localization in accelerators typically uses a small scattering target as a primary element and a bulk absorber as a secondary element. The role of the primary element is to give a substantial angular kick to the incoming particles in order to increase the impact parameter on the secondary element, which is generally placed in the optimum position to intercept transverse or longitudinal beam halos. An amorphous primary target scatters the impinging particles in all possible directions. Ideally, one would prefer to use a "smart target" which kicks all particles in only one direction: for instance, only in the radial direction, only outward, and only into the preferred angular range corresponding to the center of the absorber (to exclude escapes). A bent crystal is the practical implementation for such a smart target: it traps particles and conveys them into the desired direction. Here, the random scattering process on single atoms of an amorphous target becomes the selective and coherent scattering on atomic planes of an aligned monocrystal.

In U-70 experiment [7] different crystals in high-vacuum goniometers were serially entered in a circulating accelerated beam as shown in Fig.5. This angle of bending is sufficient to separate the circulating and deflected (by the crystal) beams in space. The beam deflection effect due to channeling was measured by secondary emission detector (SEM) located in vacuum chamber of an accelerator near to the circulating beam. Parameters of the accelerator and experimental details are described in [5].

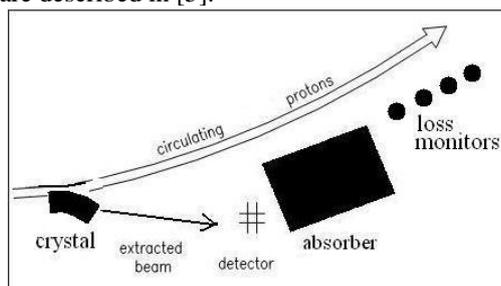

Fig. 5: The scheme of experiment on beam collimation by crystals.

The measurements have been lead at energy of protons 50 GeV. Fig. 6 illustrates the beneficial effect of crystals when used as primary element of system concerning a beam (the beam was brought to a crystal with the help of slowly increasing bump). It shows beam profiles in the radial direction 23 meters downstream of the crystal as measured on the entry face of the absorber. Two cases are reported. First, the end face of amorphous absorber is used as primary target while the crystal is kept outside of the beam envelope. As expected, the beam profile is peaked at the absorber edge. The second case corresponds to the aligned crystal: in this case the crystal channels most of the incoming particles (about 90%) into the depth of the absorber.

The deep brought of particles improves collimation, or can be used for extraction of circulating particles from the accelerator. In Fig. 7 are shown orientation curves of particle losses, measured on signals of 5 ionization chambers located in a vicinity of an absorber.

Particle losses at optimum alignment of crystals decrease in 2-3 times in comparison with disoriented crystals that correspond to calculation. Approximately in as much time intensity of muon torch behind an absorber far from the accelerator should decrease that is the important factor at achievement of high intensity of beam of circulating protons in the accelerator. In Fig. 8 effects of reduction of beam losses in the accelerator behind an absorber are shown at application of crystal in comparison with the usual one-stage scheme of beam collimation by a steel absorber.

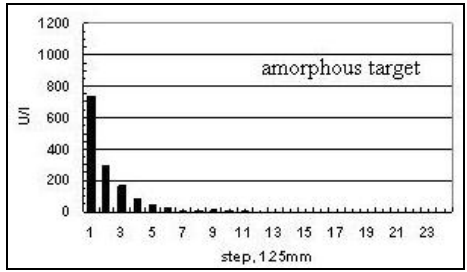 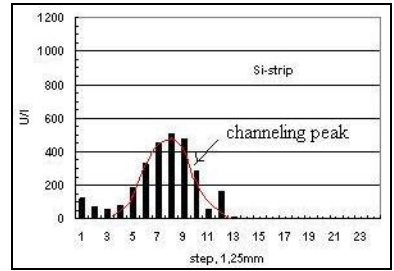

a  b

Fig. 6: Beam profiles on entry face of the absorber in different cases.

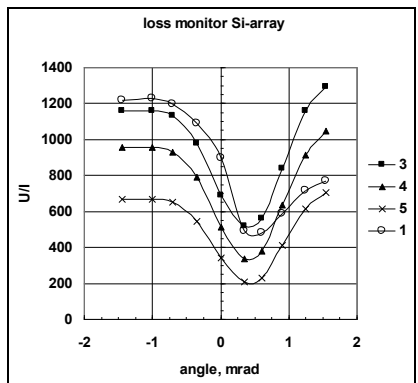 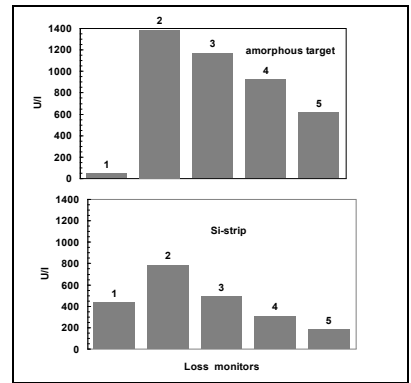

Fig. 7: Dependences of loss monitors versus orientation of crystal with respect to the beam.

Fig. 8. a - losses at usual collimation by edge of absorber, b - application of a strip – type crystal.

### 3. – Crystal extraction at low energy.

The phenomenon of deflection of a charged particle beam in a bent crystal is well investigated and successfully applied for beam extraction in high-energy accelerators at energies of about 10 GeV and higher. However, the task of bending and extraction of charged particles with energies below 1 GeV presents a big practical interest, e.g. for example for the production of ultra stable beams of low emittance for medical and biological applications. There exists a big experimental problem in steering such energy beams, which is connected with the small size of the bent crystal samples. The efficiency of particles deflection is determined by the ratio of the critical channeling angle $\theta_c$ to the beam divergence $\varphi$ and drops exponentially with the crystal length L: Eff~$(\theta_c/\varphi)\times\exp(-L/L_d)$, where the characteristic parameter $L_d$, called dechanneling length, is relatively small for low energy. For example, at E= 500 MeV we have $\theta_c$= 0.24 mrad and $L_d$= 0.4 mm. With usual channeling bent crystals (about 1 mm in length) only 10% efficiency was achieved for the deflection of sub – GeV energy particles in beam line.

Still the big problems arise in a task of extraction of a circulating beam from the ring accelerator as in addition the significant cross-section sizes of a crystal exceeding its length here are required. Thus the bend angle of a crystal should be more than 1 mrad so that the deflected beam was well separated from circulating one. Potentially suitable tools in this case can be the bent quasimosaic crystals such as in [8] or thin straight crystals [9, 10], but in both these cases it is necessary to increase a deflection angle of particles in few times. For low energy we propose a novel crystal technique, which can effectively work in a wide energy range and is especially perspective for low energy below 1 GeV.

The first option is based on use of array of shot bent channeling crystals (Fig. 9) with sub – millimeter length (special thin silicon wafers about 100 micron thickness were used for the production of such samples). Thus the bend of array occurs also, as a bend of the single well investigated silicon strip [11].

The second option is based on the reflection of particles on very thin straight crystal plates with thickness, which is equal to an odd number of half-lengths of channeling oscillation waves $L = \lambda (2n+1)/2$ where $\lambda = \pi d/\theta_c$, $d = 2.3$ $A^0$ – interplanar distance in silicon. It means for example that the optimum length of a crystal should be 10 microns for particles with energy of 50 GeV. The reflection angle in one silicon plate should be equal to twice the critical angle $\theta_c = (2U_o/pv)^{1/2}$, where: $U_o \sim 22$ eV – is the value of the potential of planar channel in silicon; p, v - a momentum and speed of incident particle. For the enhancement of deflection angle, the few aligned plates placed like a veer are foreseen (Fig. 10).

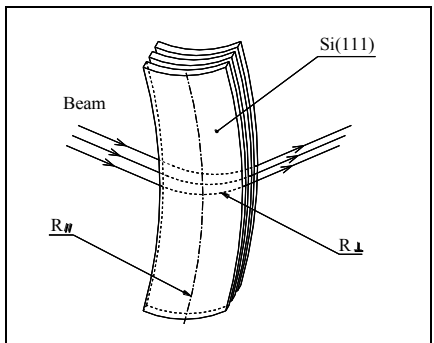 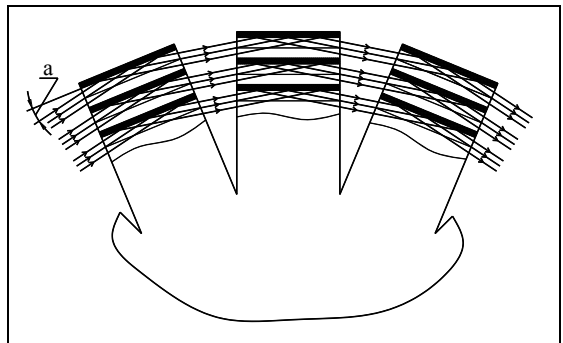

Fig. 9: The array of bent silicon strips for beam deflection due to channeling.

Fig. 10: Veer-type reflector for bending of particle beam with use of thin straight crystals. Reflection of trajectories of particles from nuclear planes is schematically shown.

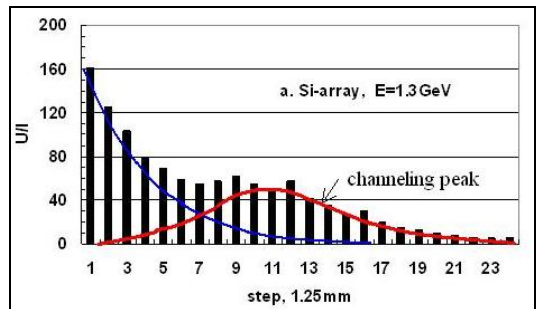

Fig. 11: The 1.3 GeV proton beam profile at the absorber entry deflected by array of silicon strips.

Recently crystal extraction/collimation experiments in U-70 are started at low energy of 1.3 GeV with new crystal technique [7]. The first result, channeling peak of about 20%, has shown array from seven thin strip crystals. The big loss of efficiency is explained by non-optimal tuning of circulating beam towards the crystal by bump-magnet. At low energy because of sizable beam, about 50 mm, there is a drift of an incident angle about a half milliradian that should be removed at prompting a beam by high-frequency noise (this work is planned). In Fig. 11 the profile of 1.3 GeV beam is shown deflected by silicon array. The fraction of channeling peak is allocated by a thick line (channeling peak is well separated from a circulating beam and approximately corresponds to efficiency of a possible beam extraction from the accelerator).

**4. – Perspectives of beam collimation on basis of reflections in axially-oriented crystals.**

Recently IHEP group together with employees of several Russian and foreign centers of science have opened the new physical phenomenon - reflection of high energy protons from the bent atomic planes of silicon crystal ([3] and ref. herein). Volume reflection is caused by interaction of incident particle with potential of the bent atomic lattice and occurs on small length in the vicinity of a tangent to the bent atomic plane, leading to deflection of a particle aside, opposite to a bend. The phenomenon of reflection occurs in wide area of angles and is more effective, than usual channeling. Therefore there are real prospects to use of volume reflection for extraction and collimation of beams in the big accelerators [12].

For real application the increase in angle of reflection in few times is required. Two ways were proposed for this purpose:
- Reflection on a chain of crystals (multiple volume reflection - MVR in sequence of crystals [13]).
- Reflection near to an axis in total potential of several skew planes (MVR in one crystal [14,15].

The first testing of multi-reflection structures with planar alignment in U-70 circulating beam shows promising results [16]. This technique was also successfully tested in the Tevatron [17].

Now in [18] we tested amplification of reflection angle due to both effects (multi-crystals and axial enhancement) for improvement of beam collimation scheme in U-70 synchrotron at 50 GeV. Multicrystal structure, 6 silicon bent strips (Fig.12a), was prepared by our technology described in [16]. This structure was installed in two-axes goniometer in U-70 circulating beam (Fig.12b) like a first stage of collimation system. Reduction of particle losses in accelerator was observed during planar crystal rotation which was increased by vertical rotation like shown in Fig.13. This effect is fully explained by Monte Carlo simulations presented in Fig. 14 where planar reflections and axial enhancement in crystal are shown.

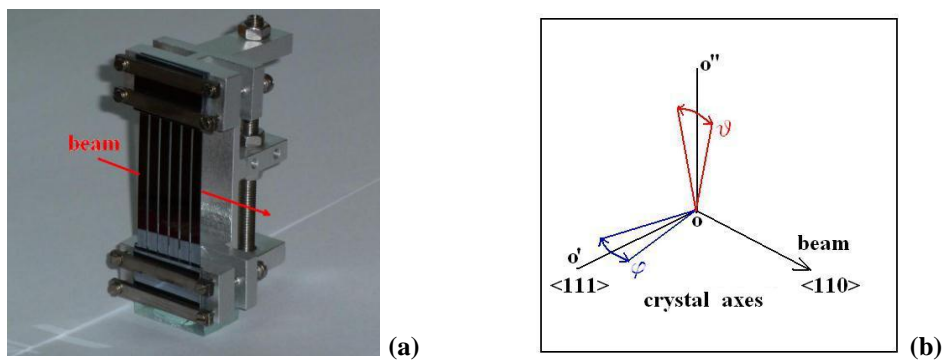

Fig.12: Crystal multistrip structure and scheme of its rotation in two-axes goniometer.

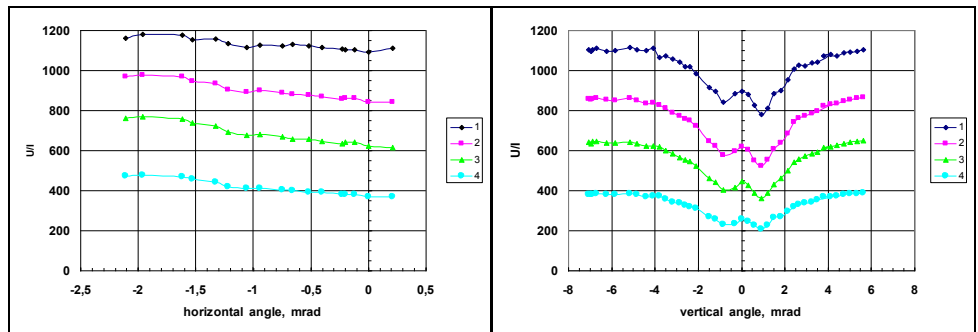

Fig.13: Reduction of particle losses of circulating beam in four places downstream collimator versus horizontal and vertical crystal rotation.

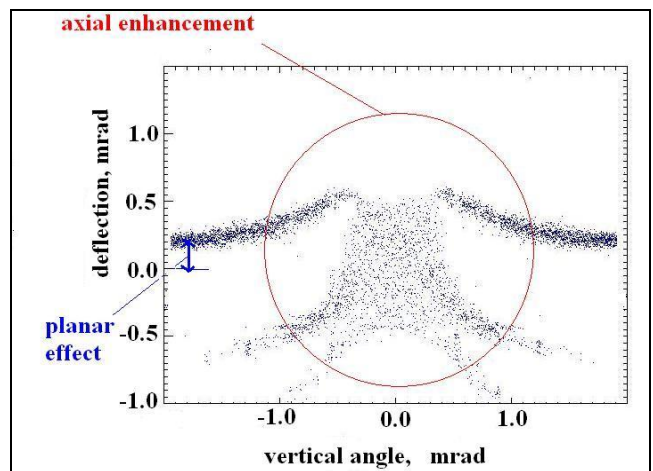

Fig.14: The dependence of 50 GeV proton deflection versus vertical crystal angle at one passage in crystal multistructure initially aligned in mode of planar reflections.

Thus the new method of beam steering was demonstrated, based on reflections of particles in multi-crystal enhanced by axial effect, which is very promising also for negative particles.

This work is supported by IHEP Directorate, State corporation Rosatom and RFBR grant 08-02-01453a.

**References.**

[1] ELISHEV A.F. et al., Phys.Lett.**B88**:387(1979), JETP Lett.**30**:442(1979).
[2] FLILLER R.P. et al., Phys.Rev.ST Accel.Beams **9**:013501, 2006.
[3] SCANDALE W. et al., Phys.Rev.Lett.**98**:154801, 2007,
    SCANDALE W. et al., Phys.Lett.**B692**:78-82, 2010.
[4] BIRYUKOV V.M., CHESNOKOV Y.A., KOTOV V.I. Crystal channeling and its
    application at high-energy accelerators. Berlin, Germany: Springer (1997) 219 pp.
[5] AFONIN A.G. et al., Phys.Part.Nucl.**36**:21-50, 2005.
[6] ARHIPENKO A.A. et al., JETP Letters, **88**(4), 265 (2008).
[7] AFONIN A.G. et al., IHEP Preprint 2010-12;
    JETP letters, **92**, 206-209, 2010.
[8] IVANOV Y.M. et al., JETP Letters **81** (2005) 99.

[9]   TARATIN A. et al., SSCL-545, 1991.
[10] STROKOV S., et al., Nucl.Instrum.Meth.**B252**:16-19, 2006.
[11] AFONIN A.G. et al., Phys. Rev.Lett. **87**, 094802 (2001).
[12] YAZYNIN I.A. et al Proc. SPIE Int.Soc.Opt.Eng.6634:66340H, 2007.
[13] SCANDALE W. et al., Physical Review Letters **102**, 084801 (2009).
[14] TIKHOMIROV V., Physics Letters, **B655** (2007) 217.
[15] SCANDALE W. et al., Physics Letters **B682** (2009), 274–277.
[16] AFONIN A.G. et al., Atomic Energy, **106**, 409-414, 2009.
[17] SHILTSEV V. et al., FERMILAB-CONF-10-127-APC, May 2010. 3pp, IPAC'10, Kyoto, Japan, 23-28 May 2010.
[18] AFONIN A.G. et al., JETP Letters **93** (2011) 187 .